\documentclass[twocolumn]{iopart}

\usepackage{graphicx}  % needed for figures
\usepackage{bm}        % for math
\usepackage{amssymb}   % for math
\usepackage{color}
\usepackage{subcaption}
\usepackage[utf8]{inputenc}
\usepackage{braket}
\usepackage{upgreek}

\usepackage{xurl}
\usepackage[linkcolor=blue,colorlinks=true,urlcolor=blue,citecolor=blue]{hyperref}
\usepackage[nameinlink]{cleveref}

\newcommand{\fm}{f_\mathrm{m}}
\newcommand{\fs}{f_\mathrm{s}}
\newcommand{\fd}{f_\mathrm{d}}
\newcommand{\fp}{f_\mathrm{p}}
\newcommand{\vp}{V_\mathrm{p}}
\newcommand{\vac}{V_\mathrm{ac}}
\newcommand{\vdc}{V_\mathrm{dc}}

\begin{document}

\title{Random number generation with a chaotic electromechanical resonator}

\author{Guilhem Madiot}
\address{Centre de Nanosciences et de Nanotechnologies, CNRS, Université Paris-Saclay, 91120 Palaiseau, France}

\author{Franck Correia}
\address{Centre de Nanosciences et de Nanotechnologies, CNRS, Université Paris-Saclay, 91120 Palaiseau, France}

\author{Sylvain Barbay}
\address{Centre de Nanosciences et de Nanotechnologies, CNRS, Université Paris-Saclay, 91120 Palaiseau, France}

\author{Remy Braive}
\address{Centre de Nanosciences et de Nanotechnologies, CNRS,  Université Paris-Saclay, Palaiseau, France}
\address{Université de Paris, F-75006 Paris, France}
\address{Institut Universitaire de France, Paris, France}
\ead{remy.braive@c2n.upsaclay.fr}

\vspace{10pt}

\begin{abstract}
Chaos enables the emergence of randomness in deterministic physical systems. Therefore it can be exploited for the conception of true random number generators (RNG) mandatory in classical cryptography applications. Meanwhile, nanomechanical oscillators, at the core of many on-board functionalities such as sensing, reveal as excellent candidates to behave chaotically. This is made possible thanks to intrinsic mechanical nonlinearities emerging at the nanoscale. Here we present a platform gathering
a nanomechanical oscillator and its integrated capacitive actuation. Using a modulation of the resonant force induced by the electrodes, we demonstrate chaotic dynamics and study how it depends on the dissipation of the system. The randomness of a binary sequence generated from a chaotic time trace is evaluated and discussed
such that the generic parameters enabling successful random number generation can be established. This demonstration makes use of concepts which are sufficiently general to be applied to the next generation of nano-electro-optomechanical systems.
\end{abstract}
\ioptwocol

\section{Introduction}

The combination of integrated electronics and suspended micro or nanomechanics in micro-nano electromechanical systems (M\&NEMS)  have led to a large number of industrial applications that have now invaded our daily lives such as, e.g., accelerometers or gyroscopes \cite{bogue2007mems}, all present in a modern smartphone. M\&NEMS are also likely to be used as gas, mass or pressure sensors, and have also potential for bio-medical applications \cite{bhansali2012mems}. Although these devices generally rely on the static response of a mechanical component to an external stimulus, e.g. the acceleration provoked by a car accident in an air-bag trigger, it can also be interesting to exploit the resonance phenomena occurring in these mechanical structures. By coupling the latter with an excitation scheme, such as a piezoelectric or capacitive actuator, one can resonantly drive the mechanical motion. This configuration finds an immediate application with microphone which convert electrical signals to acoustic waves, and reciprocally. By reducing the dimensions of these electromechanical systems at the nanoscale, one can not only access the radio-frequency (RF) domain, but also unavoidably allow nonlinear phenomena to manifest in the dynamics of these devices. Interestingly such nonlinear behavior is not necessarily a drawback but can actually be exploited. Amplification of weak signals, for example, can be achieved using NEMS thanks to a bistable regime enabled by a structural Duffing anharmonicity of the material \cite{chowdhuryPRL,chowdhury2019weak}.

Among the possible regimes achievable with such nonlinear oscillators, chaos might be the most intriguing as it enables the introduction of unpredictability in classical and deterministic physical systems. Thus chaos emerges as a possible solution to generate true randomness without appealing to stochastic \cite{gong2019true} or quantum \cite{Vallone2014} phenomena. In this spirit there have been several proposals to generate true random number sequences out of a chaotic time trace \cite{shore2015,kim201782,shi2020gbps,yu2019survey,Shi:20}. 
Several approaches using mechanical systems have also been considered \cite{Voris2011,app9010185,Dantas2020}, for pseudo-RNG applied to image encryption \cite{haliuk2022memristive}, and applied with bit-rate above 100 MHz \cite{garcia2017fast}, but rely on an external chaotic generator.

Here we present an integrated electromechanical device based on photonic-crystal (PhC) membrane and interdigitated electrodes (IDE) separated by a nanometric air-gap (see \cref{fig1:a,fig1:b}). Such a system enables controlled chaos to emerge using a slowly-modulated electromechanical force \cite{defoort2021dynamical,HouriPRL,miles1984chaotic}. An in-depth study of chaotic dynamics is performed upon the mechanical dissipation. Applying a threshold to generate random binary sequences from chaotic time traces, we evaluate the randomness quality as a function of two parameters involved in the sequence generation. Importantly, this study goes way beyond the scope of our system and could be applied to any system displaying similar driven chaotic dynamics \cite{HouriPRL}.

Our experiment, which gathers both electromechanical actuation and optical readout aided by a integrated optical cavity, enables the transduction of the mechanical motion into the optical domain. This ability to gather electromechanical and optical properties within the same device constitutes the basis for the development of nano-opto-electro-mechanical systems (NOEMS) \cite{Zobenica2017,Midolo2018,nanXu2021,navarro2022}.
The multiplication of such novel platforms respond the interest focused to specific functionalities, such as optical-to-RF conversion \cite{bochmann2013nanomechanical,Bagci2014,balram2016coherent}, or opto-electro-mechanical switches  \cite{Liu:17,Haffner2019}, for example.

\section{System description and Nanofabrication}

\begin{figure*}[!ht]
\centering
\includegraphics[scale=0.96,trim={0.4cm 0 0.8cm 0cm},clip]{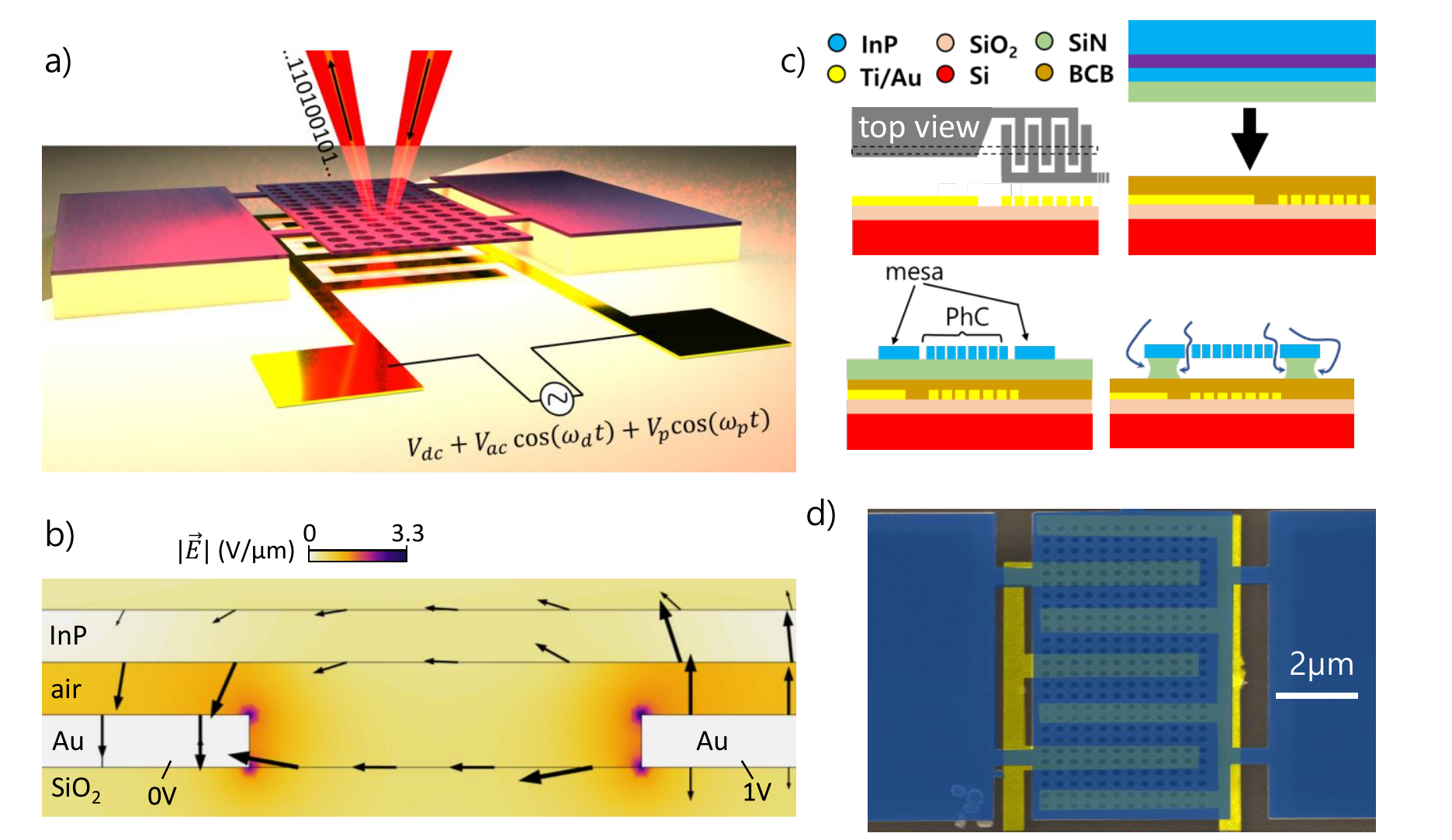}
{\phantomsubcaption\label{fig1:a}}
{\phantomsubcaption\label{fig1:b}}
{\phantomsubcaption\label{fig1:c}}
{\phantomsubcaption\label{fig1:d}}
\caption{a) Artistic view of the platform and random number generation encrypted in the readout optical field. b) FEM simulation of the electric field in a simplified cross-section of the system. The black arrows show the vector field $\overrightarrow{E}$ which its norm is shown in color. c) Fabrication process: IDEs deposition by EBL and lift off, integration of an InP substrate by heterogeneous bonding, EBL definition of the mechanical structures and ICP, under-etching and CPD. d) colorized SIM micrograph with the PhC membrane (blue) and IDT shown in transparence (yellow).}
\end{figure*}

To gather the mechanical, electrical and optical properties into such NOEMS, we base the design on a 260 nm thick suspended InP membrane. This $20\times10$ $\upmu$m$^2$ rectangular membrane constitutes a mechanical resonator whose quality factor can be significantly increased by optimizing the four bridges connecting the membrane to the substrate \cite{chowdhury2016mechanical}. Electromechanical actuation is enabled by the integration of IDE below the free-standing membrane.
The latter is engineered as an optical reflector by etching it through with a 2D photonic crystal which maximizes its normal reflexion coefficient \cite{Antoni:11}. 

%\begin{figure}[!ht]
%\centering
%\includegraphics[scale=0.5,trim={0cm 0cm 0 0cm},clip]{figs/fig2bis.pdf}
%\caption{Fabrication process a) IDTs deposition by EBL + lift off b) integration of an InP substrate by heterogeneous bonding c) EBL definition of the mechanical structures + ICP d) under-etching + CPD}
%\label{fig2}
%\end{figure} 

The main steps of the nanofabrication process flow are schematically depicted in \cref{fig1:c}. They include a) the electron beam lithography (EBL) of the IDE on a silicon wafer, followed by metal deposition and lift off; b) the heterogeneous BCB-bonding \cite{BCBref} on the Si wafer of a InP substrate incorporating the InP membrane onto which 350 nm of SiN is deposited by PECVD and chemical InP substrate removal; c) the EBL patterning of the mechanical structures and photonic crystals followed by ICP etching of the InP layer; and d) Under-etching of the mechanical structure (etching of the SiN layer) with HF, followed by critical point drying. The resulting platform is shown in the colorized SEM micrograph in \cref{fig1:d}.

The $350$ nm air gap separating the membrane from the IDE placed below constitutes an optical cavity at the He-Ne wavelength (633 nm).
This optomechanical readout permits to enhance the sensitivity to detect the membrane out-of-plane resonances \cite{MadiotPRA2021}. 
Meanwhile the dielectric properties of the membrane makes it easier to be excited by capacitive actuation. Both DC gate $\vdc$ and RF signal $\vac\cos(2\pi \fd t)$ are applied externally between these two electrodes and the field lines penetrate and polarize the InP membrane. In \cref{fig1:b}, we show the stationary solution for the electric vector field $\overrightarrow{E}$ (arrows) and norm $|\overrightarrow{E}|$ (colorscale) in the system, obtained by Finite-Element Simulation methods. This model accounts for a 2D cross-section of the structure, with electrical potential of 1V imposed every two gold digit while the others are grounded. This produces electric field arcs that cross the above InP membrane, which induces a capacitive force on it. Experimentally, the oscillating contribution of this force is proportional to $\vdc\vac\cos(2\pi \fd t)$.
Therefore when the actuation frequency $\fd$ approaches a mechanical resonance $\fm$, the mechanical oscillation amplitude of the membrane increases. 
With this system, we are interested in probing the first order mode of the membrane which has the highest overlap with the incident optical field. This "drum" mode has the strongest out-of-plane amplitude and therefore is prone to exhibit nonlinearity, which is essential in the following to obtain a chaotic dynamics.

%\begin{figure}[!ht]
%\centering
%\includegraphics[scale=1]{figs/fig3.pdf}
%{\phantomsubcaption\label{fig3:a}}
%{\phantomsubcaption\label{fig3:b}}
%{\phantomsubcaption\label{fig3:c}}
%\caption{a) Design the IDE. Units are in $\upmu$m, a=1$\upmu$m. b) SEM micrograph of the IDE, before III-V integration. c) Colorized SEM micrograph of the final system. Gold IDE are shown in yellow, InP membrane is blue.}
%\end{figure}

%By mechanically connecting two identical membranes with a $1.5\times0.5$ $\upmu$m$^2$ junction, we obtain a system of two-coupled resonators with independent electromechanical actuations (see \cref{fig3:b,fig3:c}). Such mechanical coupling leads to the hybridization of the two drum-modes of the membranes \cite{MadiotPRA2021,Doster2019,Okamoto2013}.
%In what follows, we focus on one normal mode of the two-membranes system, namely the symmetric mode, i.e. associated to in-phase motion of the membranes. Provided that the spectral splitting between the two normal modes ($\sim208$ kHz) is much bigger than the mechanical linewidth, we can model this mode as a single resonator.

\section{Dissipation parameters control and influence on chaotic dynamics}

\begin{figure}[!ht]
\centering
\includegraphics[scale=0.96]{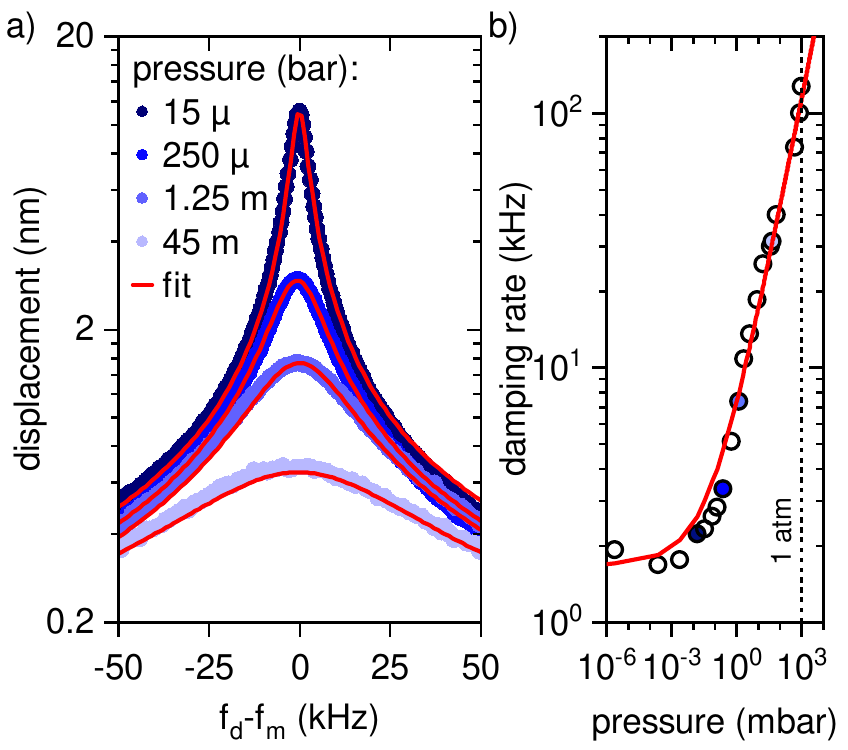}
{\phantomsubcaption\label{fig4:a}}
{\phantomsubcaption\label{fig4:b}}
\caption{a) Normalized spectral response of the mechanical mode at different pressure condition in the vacuum chamber with Lorentzian fits (red lines). b) Pressure dependence of the mechanical damping (do, with a power-law fit (red line).}
\end{figure}

The sample is placed in a vacuum chamber pumped at $10^{-6}$ mbar. The He-Ne laser is focused at the center of the membrane. Several mechanical modes are observed between 2 and 15 MHz.
As discussed above, we focus on the fundamental mode, with frequency $\fm=2.327$ MHz. The associated IDE is submitted to a voltage $\vdc + \vac\cos(2\pi\fd t)$ with $\vdc=2$ V and $\vac=0.5$ V.
While scanning the driving frequency $\fd$ around the mechanical resonance, we demodulate the photoreceived optical signal with a passband filter centered at $\fd$ and 100 Hz wide, returning both demodulated amplitude $V$ and phase $\varphi$. The amplitude voltage can be converted into a mechanical displacement $r$ after calibration. \

This measurement is reproduced for different values of the pressure in the vacuum chamber and four representative spectral responses are shown in \cref{fig4:a}. 
We note a significant broadening of the mechanical resonance while the pressure increases. This results from the increasing contribution of air-damping in the mechanical dissipation. 
We fit each resonance curve with the lorentzian lineshape (red lines) and extract the mechanical damping rate. The latter is plotted as a function of the pressure in \cref{fig4:b}. An exponential increase is observed above 1 mbar while a saturation to $\gamma_0\sim2$ kHz is observed below this value. The saturation arises when the mechanical losses are no longer dominated  by air-damping but rather explainable by internal mechanical loss channels such as clamping and thermomechanical losses.

\begin{figure*}[!ht]
\centering
\includegraphics[scale=1]{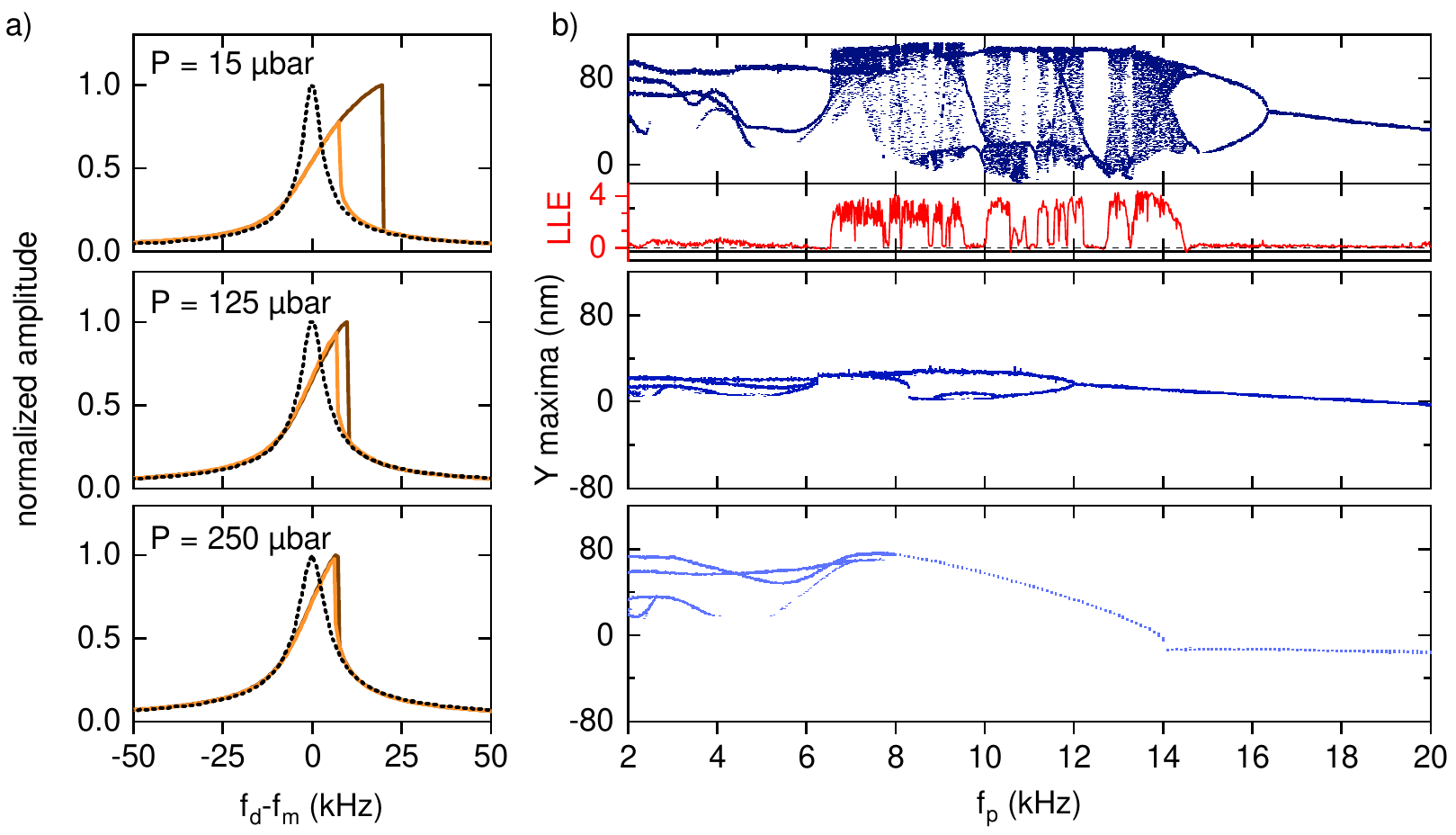}
{\phantomsubcaption\label{fig5:a}}
{\phantomsubcaption\label{fig5:b}}
{\phantomsubcaption\label{fig5:c}}
\caption{For 3 different values of the pressure in the vacuum chamber : a) Normalized mechanical displacement with forward (brown) and backward (orange) sweep of the driving frequency around mechanical frequency $\fm=2.327$ MHz. The linear response is indicated for reference (back dashed). b) Experimental bifurcation diagram parametrized with the pump frequency $\fp$.}
\end{figure*}

So far we have used sufficiently low input voltages $\vac$ and $\vdc$ for the mechanics to remain in the linear regime, where the system can be treated as a driven harmonic oscillator. Increasing the applied electromechanical force leads to nonlinear responses that typically manifest itself by a mechanical bistability.
This is experimentally verified by scanning $\fd$ forward and backward around the mechanical frequency. 
Such measurement is shown in \cref{fig5:a} at three different values of the vacuum chamber pressure. The data are normalized by their respective maximum and the linear lineshape (see \cref{fig4:a}) are reported here for reference. In each case we apply the same excitation with $\vac=3$ V. We note that a relatively small increase of the mechanical linewidth provokes a significant change in the bistability, whose frequency span tends to reduce. 

In the following we induce a chaotic dynamics by performing an amplitude modulation of the electromechanical force \cite{MadiotPRA2021,defoort2021dynamical,HouriPRL,miles1984chaotic}. We add a new RF drive -- we refer as the pump signal -- $\vp\cos(2\pi\fp t)$ which leads to near-resonant force amplitude $F(t)\propto\vdc\Big(\vac+\vp\cos(2\pi\fp t)\Big)$. The driving frequency $\fd$  is set at the low-frequency edge of the bistability, ensuring a wider bandwidth for the chaotic regime \cite{defoort2021dynamical}. 

Both the pump amplitude $\vp$ and frequency $\fp$ can be played with to tune the dynamical regime of mechanical responses. Increasing the amplitude for a fixed frequency typically leads to a period-doubling cascade route to chaos \cite{MadiotPRA2021}. Here we rather set the pump amplitude to $\vp=2.5$ V and scan the frequency from 2 kHz to 20 kHz. The mechanical response time traces $r(t)$ and $\varphi(t)$ are recorded for 100 ms. The quadrature $Y(t)=r(t)\sin\big(\varphi(t)\big)$ is used to reconstruct a Poincaré section of the signal. The latter is plotted as a function of $\fp$ in the bifurcation diagrams shown in \cref{fig5:b}.

At low pressure, the mechanical displacement induces sufficiently strong nonlinearity to enable a driven chaotic dynamics, as illustrated at $P=15$ $\upmu$bar. The cascaded period doubling to chaos is observed for decreasing $\fp$, and starts around 16.5 kHz. The presence of chaos can be numerically verified by computing the Largest Lyapunov Exponent, that we show in red below this bifurcation diagram. A zero LLE indicates a periodic or quasi periodic motion whereas a strictly positive LLE corresponds to a chaotic dynamics. The presence of experimental noise slightly increases the LLE giving rise to a slighly positive value even for periodic motion but the significant increase at some specific positions -- from 6.6 to 14.5 kHz -- clearly indicates the range of $\fp$ in which chaos emerges. These regions are also visually identifiable on the diagram as they translates into a dense Poincaré section. Here the chaos spans over $\sim8$ kHz but this range tends to reduce when the mode linewidth increases, since this comes also with a decrease of the mechanical nonlinearity, the bistability span being strongly dependent on the damping rate.
Thus at $125$ $\upmu$bar the bifurcation diagram does not display a chaotic regime although a period-doubling is observed at $\sim12$ kHz. The dynamics keeps getting poorer as the pressure in the chamber increases, as illustrated with the third panel taken at 250 $\upmu$bar. 

%-----------------------------------------

\section{Random number generation}

\begin{figure}[!ht]
\centering
\includegraphics[scale=1]{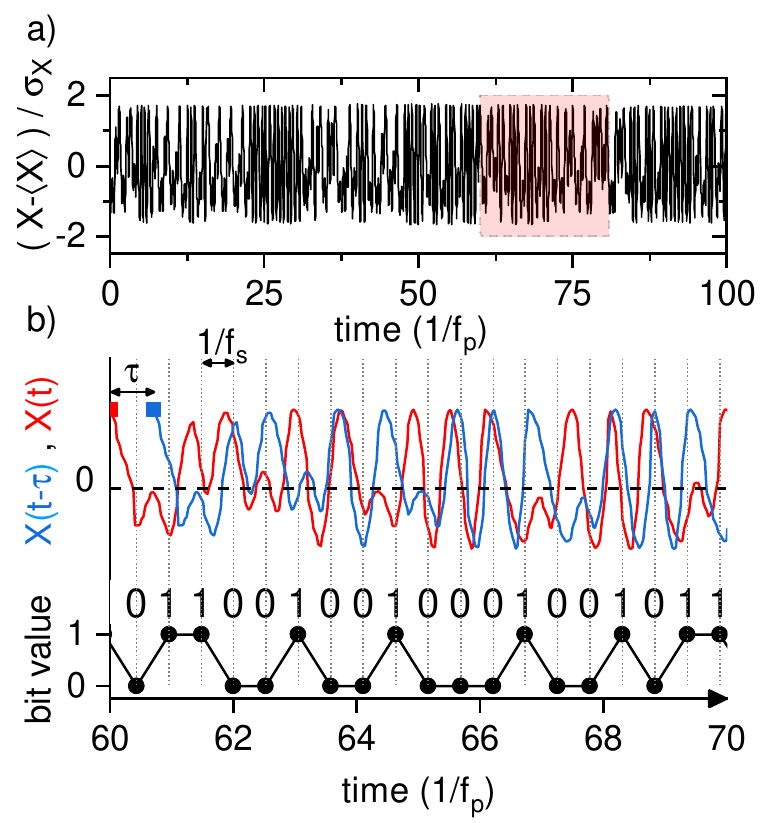}
{\phantomsubcaption\label{fig6:a}}
{\phantomsubcaption\label{fig6:b}}
{\phantomsubcaption\label{fig6:c}}
\caption{a) Experimental chaotic time trace. b) The trace $X(t)$ is compared to its delayed self $X(t-\tau)$. c) The XOR logical gate is periodically applied to the traces relative sign. It results in a binary sequence. }
\end{figure}

Now that we have described the conditions under which chaos can be reached, we can focus on the exploitation of this dynamical regime to generate random numbers. Here we apply a method enabling the production of a sequence of random bits from an experimental chaotic time trace. This is achieved using a method described in \cite{Hirano:10} where the sign of a time trace is periodically compared to a delayed copy of itself with the logical function XOR. We record the quadrature $X(t)=r(t)\cos\big(\varphi(t)\big)$ for $\fp=12.3$ kHz and $\vp=2.5$ V. In practice, we consider the normalized time trace $(X-\braket{X})/\sigma_X$ where $\braket{X}$ and $\sigma_X$ are the mean value and the standard deviation of $X(t)$ calculated over the full time trace, respectively. Note that the following results remain unchanged by using the other quadrature, $Y(t)$. The normalized time trace under study is shown in \cref{fig6:a}. We introduce a delay $\tau$ and a sampling frequency $\fs$, which corresponds to the rate at which $X(t)$ and $X(t-\tau)$ will be compared (see \cref{fig6:b}). The bits resulting from the XOR gate applied between the respective signs of these two traces are shown in \cref{fig6:c}.

In the following the randomness of a binary sequence is verified by applying the NIST Statistical Test Suite \cite{NIST_StatTestSuite}. It is composed by 14 randomness tests, each returning a p-value that can be interpreted as the probability for the sequence to be random according to the corresponding test. The p-value validates the test if its value is above $0.01$. In the following we simply apply all these algorithms on our binary sequence and check the p-values. If all the p-values validate the sequence as random, we consider that this sequence passes the randomness test. On the contrary, if at least one test fails, we consider the sequence as not random.

\begin{figure}[!ht]
\centering
\includegraphics[scale=1]{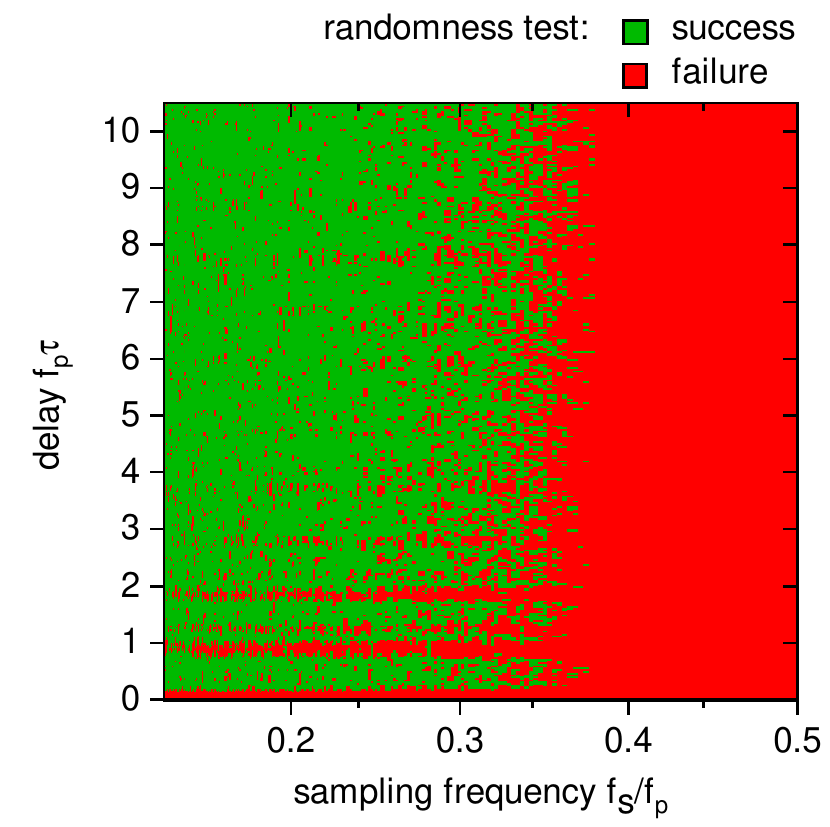}
\caption{Randomness test suite result as a function of the delay $\tau$ and the sampling frequency $f_s$ applied to the trace presented in \cref{fig6:a}. Both quantities are shown in units of modulation period.}
\label{fig7}
\end{figure}

Our objective is to characterize the randomness test success as a function of the delay $\tau $ and the sampling frequency $f_s$. By generating a binary sequence for several values of $\tau$ and $f_s$, we plot a matrix showing the randomness test result in \cref{fig7}. The green (resp. red) pixels correspond to a successful (resp. unsuccessful) test. Both the the sampling frequency and the delay are shown in units of pump frequency and pump period, respectively. $\fp^{-1}$ corresponds to the mean oscillation period of the chaotic trace.
We observe an significant increase of the randomness quality towards low sampling frequencies, with a threshold limit around 0.35. %One can interpret this value as a limit above which the time left to the signal for randomness to emerge would not be sufficient. %This is shown on top of the map where the mean success is calculated over all values of the delay (from 0 to $\fp$) and plotted as a function of $\fs$. We fit an exponential in orange.
This is related to the modulation frequency $\fp$ in the present case. Indeed at high sampling frequencies, the trace does not have enough time to evolve between two samples. This favors the apparition of runs of ones and zeros ('00','11','000','111', and so on) which break the sequence randomness.
Moreover the randomness quality shows significant degradation at $\fp\uptau=1$ and $2$. This relates with a strong correlation between $X(t)$ and $X(t-\uptau)$ for these values of the delay. 
Overall, low sampling frequency and high delay improve the randomness of the sequence. Randomness can emerge only if $\tau$ is sufficiently high for the two trace to decorrelate. %Additionally, the sampling rate must preferably be chosen as lower than the pump frequency, $\fp$.
Beyond these specific points, \cref{fig7} evidences many pairs of delay and sampling frequency allowing random number generation.

\section{Conclusion}

We have presented a nano-electromechanical system displaying driven chaotic dynamics when submitted to a periodically time-varying force. In a first experiment, we submit the mechanical system to different environmental condition, by controlling the pressure in the vacuum chamber. 
From the dissipative properties of the system, we conclude on their effect on the dynamics of the nonlinear system. The emergence of chaos is clearly favoured by a higher mechanical quality factor, as this comes with a more pronounced bistability.

Using a chaotic time trace to generate binary sequences, we study the randomness of the latter as a function of two control parameter. We draw general conclusions about the respective influence of these two buttons, and relate these dependencies with the physics of the system. RNG in a driven chaotic system such as the one presented here could benefit from the use of two quadratures while only one is considered here. Simultaneous generation of random bits using both quadratures could multiply the bit-rate by two, and even more in multimode systems, if several drives and demodulation channels are used at the same time \cite{MadiotPRA2021}.

In addition, the dynamics under study here is scalable, i.e. it could be reproduced in other types of devices, performing at higher frequencies, and relying of a different physics. The condition for chaos to emerge relies on the presence of an intrinsic nonlinearity triggering bistable behaviours in the system on the one hand, and on the time-modulation of the driving force on the other hand. Therefore, Kerr optical cavities as well as optomechanical system could be exploited to reproduce these results a higher frequency, in which case the subsequent bit-rate could reach several Gbs.

\ack
This work is supported by the French RENATECH network, the European Union’s Horizon 2020 research innovation program under grant agreement No 732894 (FET Proactive HOT), the Agence Nationale de la Recherche as part of the “Investissements d’Avenir” program (Labex NanoSaclay, ANR-10-LABX-0035) with the flagship project MaCaCQu and the JCJC project ADOR (ANR-19-CE24-0011-01).

\vspace{1cm}
%\appendix
%\section{Ap.A}
%\section{Ap.B}

\bibliographystyle{unsrt}
\bibliography{ms}

\end{document}